\documentclass[reprint,amsmath,amssymb,aps]{revtex4-1}
\usepackage{bm}
\usepackage{dcolumn}
\usepackage{lineno}
\usepackage{graphicx}
\usepackage{subfigure}
\usepackage{epstopdf}
\usepackage{float}
\usepackage{threeparttable}
\usepackage{overpic}
\usepackage{subfigure}
\usepackage{float}
\usepackage{ulem}
\usepackage{cmap} 
\usepackage{mathtext}
\usepackage{mathrsfs}
\usepackage{multirow}
\usepackage{xcolor}
\definecolor{aa}{RGB}{0,0,139}
\usepackage[bookmarksnumbered=true,colorlinks,urlcolor=aa,linkcolor=blue,anchorcolor=aa,citecolor=aa]{hyperref}

\newcommand{\SSB}{\Sigma^+\bar{\Sigma}^-}

\newcommand{\kk}{K^+K^-}
\newcommand{\ppb}{p\bar{p}}

\newcommand{\jpsi}{J/\psi}

\newcommand{\bfg}{\begin{figure}}
\newcommand{\efg}{\end{figure}}
\newcommand{\bitm}{\begin{itemize}}
\newcommand{\eitm}{\end{itemize}}
\newcommand{\bnum}{\begin{enumerate}}
\newcommand{\enum}{\end{enumerate}}
\newcommand{\btbl}{\begin{table}}
\newcommand{\etbl}{\end{table}}
\newcommand{\btbu}{\begin{tabular}}
\newcommand{\etbu}{\end{tabular}}
\newcommand{\bcl}{\begin{center}}
\newcommand{\ecl}{\end{center}}

\newcommand{\beq}{\begin{equation}}
\newcommand{\eeq}{\end{equation}}
\newcommand{\beqr}{\begin{eqnarray}}
\newcommand{\eeqr}{\end{eqnarray}}

\definecolor{boslv}{rgb}{0.0, 0.65, 0.58}
\definecolor{Munsell}{HTML}{00A877}

\newcommand{\Br}{\mathcal{B}}

\begin{document}
\title{Observation of the $\psi(3686)$ decays into $\Sigma^{+}\bar{\Sigma}^{-}\omega$ and $\Sigma^{+}\bar{\Sigma}^{-}\phi$}

\author{
\begin{small}
\begin{center}
M.~Ablikim$^{1}$, M.~N.~Achasov$^{5,b}$, P.~Adlarson$^{75}$, X.~C.~Ai$^{81}$, R.~Aliberti$^{36}$, A.~Amoroso$^{74A,74C}$, M.~R.~An$^{40}$, Q.~An$^{71,58}$, Y.~Bai$^{57}$, O.~Bakina$^{37}$
% ...
\\
\vspace{0.2cm}
(BESIII Collaboration)\\
\vspace{0.2cm} {\it
$^{1}$ Institute of High Energy Physics, Beijing 100049, People's Republic of China\\
% ...
$^{5}$ Budker Institute of Nuclear Physics SB RAS (BINP), Novosibirsk 630090, Russia\\
% ...
$^{36}$ Johannes Gutenberg University of Mainz, Johann-Joachim-Becher-Weg 45, D-55099 Mainz, Germany\\
$^{37}$ Joint Institute for Nuclear Research, 141980 Dubna, Moscow region, Russia\\
% ...
$^{40}$ Liaoning Normal University, Dalian 116029, People's Republic of China\\
% ...
$^{57}$ Southeast University, Nanjing 211100, People's Republic of China\\
$^{58}$ State Key Laboratory of Particle Detection and Electronics, Beijing 100049, Hefei 230026, People's Republic of China\\
% ...
$^{71}$ University of Science and Technology of China, Hefei 230026, People's Republic of China\\
% ...
$^{74}$ University of Turin and INFN, (A)University of Turin, I-10125, Turin, Italy; (B)University of Eastern Piedmont, I-15121, Alessandria, Italy; (C)INFN, I-10125, Turin, Italy\\
$^{75}$ Uppsala University, Box 516, SE-75120 Uppsala, Sweden\\
% ...
$^{81}$ Zhengzhou University, Zhengzhou 450001, People's Republic of China\\
\vspace{0.2cm}
% ...
$^{b}$ Also at the Novosibirsk State University, Novosibirsk, 630090, Russia\\
% ...
}\end{center}
\vspace{0.4cm}
\end{small}
}

\date{\today}


	\begin{abstract}
		Based on $(27.08\pm 0.14)\times10^{8}$ $\psi(3686)$ events collected with the BESIII detector operating at the BEPCII collider, the $\psi(3686)\to\Sigma^{+}\bar{\Sigma}^{-}\omega$ and $\Sigma^{+}\bar{\Sigma}^{-}\phi$ decays are observed for the first time with statistical significances of  13.8$\sigma$ and 7.6$\sigma$, respectively. The corresponding branching fractions are measured to be  $\Br(\psi(3686)\to\Sigma^{+}\bar{\Sigma}^{-}\omega)=(1.90 \pm 0.18 \pm 0.21) \times 10^{-5}$ and $\Br(\psi(3686)\to\Sigma^{+}\bar{\Sigma}^{-}\phi)=(2.96 \pm 0.54 \pm 0.41) \times 10^{-6}$, where the first uncertainties are statistical and the second systematic.
	\end{abstract}
	\maketitle

   \section{Introduction}\label{sec:introduction}\vspace{-0.3cm}
   Quantum Chromodynamics (QCD) is the theory of strong interactions and it has been rigorously tested at high energies~\cite{Theory_a}; in the low-energy regime, the non-Abelian nature of the theory requires
a non-perturbative approach which must rely either on lattice QCD (LQCD) or on QCD-inspired models. Therefore, experimental measurements in the low-energy regime are crucial to verify the pertinent models, calibrate their parameters, and stimulate the development of new theoretical computations~\cite{Theory_b}. As a result, many interesting properties associated with the strong decays of $J/\psi$ and $\psi(3686)$ mesons have been investigated, with the aim to advance our knowledge about the QCD in the interplay between perturbative and non-perturbative strong interaction regime~\cite{Theory_c}.
BESIII has collected large data samples of vector charmonia from $e^{+}e^{-}$ annihilations, such as $J/\psi$ and $\psi(3686)$, which provide us the opportunity to conduct detailed experimental studies on rare decays of these states and further explore the intermediate structures inherent in these processes~\cite{BESIII:Yellow_Book}.

   Among the hadronic decays, the processes of $\psi(3686)$ and $J/\psi$  decaying into baryon pairs have been understood in terms of $c\bar{c}$ annihilations into three gluons or into a virtual photon~\cite{Zhu:2015bha}. Three-body decays, e.g. $\psi(3686)\rightarrow\Lambda\bar{\Lambda}P(V)$, where $P$ represents a pseudoscalar meson such as $\pi^0$ and $\eta$, and $V$ denotes a vector meson, such as $\phi$ and $\omega$, are of great interest since they allow to investigate the intermediate excited states~\cite{theory_review}. So far, the BESIII Collaboration has reported relevant studies on the decays $J/\psi$($\psi(3686))\to\Sigma^{+}\bar{\Sigma}^{-}\eta$~\cite{xiaohao_SSP}, $\psi(3686)\to\Lambda\bar{\Lambda}\pi^{0}(\eta)$~\cite{wangshi_LLP}, and $\psi(3686)\to\Lambda\bar{\Lambda}\omega$~\cite{zhh_LLomg}, while the similar isospin-allowed decay $\psi(3686)\to\Sigma^{+}\bar{\Sigma}^{-}\omega(\phi)$ has not yet been measured. In addition, the excitation spectra of most of the hyperons are still not well understood~\cite{Sarantsev:2019xxm}. Furthermore, an enhancement around the $\Lambda\bar{\Lambda}$ production threshold was observed in the $e^{+}e^{-}\to\phi\Lambda\bar{\Lambda}$ process~\cite{enhancement_a}, but the interpretation of the $\Lambda\bar{\Lambda}$ system as originating from the $\eta(2225)$ decay, as predicted by Ref.~\cite{enhancement_b}, was rejected with a significance of $7\sigma$. A similar structure was also reported in the $B$ meson decays $B^{0}\to\Lambda\bar{\Lambda}K^{0}$ and $B^{+}\to\Lambda\bar{\Lambda}K^{+}$~\cite{enhancement_c}. Therefore, the decays of $\psi(3686)\to\Sigma^{+}\bar{\Sigma}^{-}\omega(\phi)$ provide a good opportunity to search for potential $\Sigma$ excitations and the enhancement around the $\Sigma^{+}\bar{\Sigma}^{-}$ mass threshold.

   In this paper, we report the first observations and branching fraction (BF) measurements of $\psi(3686)\to\Sigma^{+}\bar{\Sigma}^{-}\omega$ and $\Sigma^{+}\bar{\Sigma}^{-}\phi$, based on $(27.08\pm0.14)\times10^{8}$ $\psi(3686)$ events~\cite{psip_num_0912} collected with the BESIII detector. Furthermore, we search for potential excited baryon states and unknown structures in the $\Sigma\omega(\phi)$ and $\Sigma^{+}\bar{\Sigma}^{-}$ invariant mass spectra. Hereafter, we denote $\psi(3686)\to \Sigma^+\bar {\Sigma}^-\omega$ and $\Sigma^+\bar{\Sigma}^-\phi $ as $\omega$-mode and $\phi$-mode, respectively.

{\section{BESIII Detector and Monte Carlo Simulation}}

  The BESIII detector~\cite{Ablikim:2009aa} records $e^+ e^-$ collisions provided by the BEPCII storage ring~\cite{CXYu_bes3} in the center-of-mass energy range from 2.0 to 4.95~GeV,
with a peak luminosity of $1 \times 10^{33}\;\text{cm}^{-2}\text{s}^{-1}$
achieved at $\sqrt{s} = 3.77\;\text{GeV}$.
BESIII has collected large data samples in this energy region~\cite{Ablikim:2019hff,EcmsMea,EventFilter}. The cylindrical core of the BESIII detector covers 93\% of the full solid angle and consists of a helium-based multilayer drift chamber (MDC), a plastic scintillator time-of-flight system (TOF), and a CsI(Tl) electromagnetic calorimeter (EMC), which are all enclosed in a superconducting solenoidal magnet providing a 1.0~T magnetic field. The magnet is supported
  by an octagonal flux-return yoke with modules of resistive
  plate muon counters (MUC) interleaved with steel. The charged-particle momentum resolution at 1~GeV/c is 0.5\%, and the  d$E/$d$x$ resolution is 6\% for the electrons from Bhabha scattering at 1~GeV. The EMC measures photon energy with a resolution of 2.5\% (5\%) at 1~GeV in the barrel (end-cap) region. The time resolution of the TOF barrel part is 68~ps, while that of the end-cap part is 110~ps. The end-cap TOF system was upgraded in 2015 using multi-gap resistive plate chamber technology, providing a time resolution of 60~ps, which benefits $\sim83\%$ of the data used in this analysis~\cite{tof_a,tof_b,tof_c}.

   Monte Carlo (MC) simulated data samples produced with a {\sc geant4}-based~\cite{geant4} software package, which includes the geometric description~\cite{detvis} of the BESIII detector and the detector response, are used to optimize the event selection criteria, estimate the signal efficiency and the level of background. The simulation models the beam energy spread and initial-state radiation in the $e^+e^-$ annihilation using the generator {\sc kkmc}~\cite{kkmc_a,kkmc_b}. The inclusive MC sample includes the production of the $\psi(3686)$ resonance, the initial-state radiation production of the $J/\psi$ meson, and the continuum processes incorporated in {\sc kkmc}. Particle decays are generated by {\sc evtgen}~\cite{evtgen_a,evtgen_b} for the known decay modes with BFs taken from the Particle Data Group  (PDG)~\cite{pdg2022} and {\sc lundcharm}~\cite{lundcharm_a,lundcharm_b} for the unknown ones. Final-state radiation from charged final-state particles is included using the {\sc photos} package~\cite{photos}.

   To determine the detection efficiency for each signal process, signal MC samples are generated with a modified data-driven generator BODY3~\cite{evtgen_a,evtgen_b}, to simulate contributions from different intermediate states in data for a given three-body final state, as discussed in Sec.~\ref{Sec:BR_determined}.


   \section{\label{Sec:Selection}Event Selection}
   In the analysis of the $\omega$-mode, the $\Sigma^+$($\bar{\Sigma}^-$) and $\omega$ particles are reconstructed via the $\Sigma^+ (\bar{\Sigma}^-) \rightarrow p\pi^0 (\bar{p}\pi^0)$ and $\omega \rightarrow \pi^+ \pi^- \pi^0$ decays. Within these processes, the $\pi^0$ is reconstructed through the $\pi^0 \rightarrow \gamma \gamma$ decay. For the $\phi$-mode, to improve the detection efficiency, a partial reconstruction of $\psi(3686)\to\Sigma^{+}\bar{\Sigma}^{-}\phi\to p\pi^{0}\bar{p}\pi^{0}K^{+}K^{-}$ is performed, where only one $\pi^0$ is required to be reconstructed while the other  $\pi^0$ is treated as a missing particle.
	
    All charged tracks are required to satisfy $\left| {\cos \theta } \right| < 0.93$, where $\theta$ is the polar angle defined with respect to the $z$ axis, which is the symmetry axis of the MDC.
   The charged tracks not originating from $\Sigma^{+}(\bar{\Sigma}^{-})$ decays are required to have their point of closest approach to the interaction point (IP) within 10 cm along the $z$-axis, and within 1 cm in the transverse plane.
   For the charged tracks from the $\Sigma^{+}(\bar{\Sigma}^{-})$ decays, the distance of closest approach to the IP must be less than 2 cm in the transverse plane, due to the hyperons' lifetime. 
    The measurements of the flight time in the TOF and $dE/dx$ in the MDC for each charged track are combined to compute particle identification (PID) confidence levels for the pion, kaon, and proton hypotheses. The tracks are assigned to the particle type with the highest confidence level.

   Photon candidates are identified using showers in the EMC. The deposited energy of each shower must be greater than 25 MeV in the barrel region ($|\cos\theta|<0.80$) or greater than 50 MeV in the end cap region  ($0.86<|\cos\theta|<0.92$). To suppress electronic noise and energy depositions not associated with the event, the EMC cluster timing from the reconstructed event start time is further required to satisfy $0\leq t\leq$700 ns. The invariant mass of the $\pi^{0}$ candidates reconstructed from a $\gamma\gamma$ pair is constrained to the $\pi^{0}$ known mass~\cite{pdg2022} by a kinematic fit, and the $\chi^2_{\rm 1C}$ is further required to be less than 20.

   For the $\omega$-mode, in order to suppress the potential backgrounds and improve the mass resolution, a seven-constraint (7C) kinematic fit is performed for the $\psi(3686)\to p\pi^{0}\bar{p}\pi^{0}\pi^{+}\pi^{-}\pi^{0}$ hypothesis by enforcing energy-momentum conservation and constraining the invariant mass of each of the three photon pairs to the nominal $\pi^{0}$ mass. If there are more than one combination in the event, the one with the smallest $\chi_{\rm 7C}^{2}$ is chosen. Furthermore, the selection $\chi^{2}_{\rm 7C}<$45 is applied, by optimizing the figure-of-merit (FOM) defined as ${S \over {\sqrt {S + B} }}$, where $S$ denotes the number of signal events obtained from the MC simulation, while $B$ is the number of background events obtained from the inclusive MC sample.
   For the $\phi$-mode, since we reconstruct only one $\pi^{0}$ meson to enhance the reconstruction efficiency, a two-constraint (2C) kinematic fit is performed on the $p\bar{p}K^{+}K^{-}\pi^{0}$ combinations, requiring that the missing mass corresponds to the nominal $\pi^{0}$ mass, as well as the invariant mass of the photon pair. If there is more than one combination in the event, the one with the smallest $\chi^{2}_{\rm 2C}$ is chosen. Furthermore, $\chi_{\rm 2C}^2<$ 20 is required. The $\SSB$ pair candidates are selected among different $\ppb\pi^0\pi^0$ combinations by minimizing $\delta_{\omega}$$=\sqrt{(M_{p\pi^{0}}-m_{\Sigma^{+}})^{2}+(M_{\bar{p}\pi^{0}}-m_{\bar{\Sigma}^{-}})^{2}+(M_{\pi^{+}\pi^{-}\pi^{0}}-m_{\omega})^{2}}$ for the $\omega$-mode, and $\delta_{\phi}$$=\sqrt{(M_{p\pi^{0}}-m_{\Sigma^{+}})^{2}+(M_{\bar{p}\pi^{0}}-m_{\bar{\Sigma}^{-}})^{2}}$ for the $\phi$-mode, respectively.


   The $J/\psi$-related backgrounds are vetoed by requiring the $\pi^{+}\pi^{-}$($\pi^{0}\pi^{0}$) recoil mass  outside the $\jpsi$ mass window, and the $\eta$-related backgrounds are suppressed by requiring the invariant mass of $\pi^{0}\pi^{0}\pi^{0}$ outside the $\eta$ mass window. All the mass windows are determined according to their mass resolutions and are listed in Table~\ref{list:mass_windows}, where $M$ and $m$ denote the invariant mass and the known mass~\cite{pdg2022}, respectively, and $RM$ denotes the recoil mass. For the $\omega$-mode, to suppress the background channels with a number of photons different from six, the $\chi _{4C}^2\left( {6\gamma p\bar p{\pi ^ + }{\pi ^ - }} \right)$$<$$\chi _{4C}^2\left( {5\gamma p\bar p{\pi ^ + }{\pi ^ - }} \right) $ and  $\chi _{4C}^2\left( {6\gamma p\bar p{\pi ^ + }{\pi ^ - }} \right)$$<$$\chi _{4C}^2\left( {7\gamma p\bar p{\pi ^ + }{\pi ^ - }} \right) $ requirements are also added.

\begin {table}[h]
\begin{center}
  \fontsize{8}{10}\selectfont

\renewcommand\arraystretch{1.2}

    {\caption {Mass selection windows for each mode.}
    \label{list:mass_windows}}
\begin{tabular}{c|  c}
  \hline \hline

  Mode & Mass window [MeV/$c^{2}$] \\   \hline
  \multirow{6}*{\normalsize$\omega$-mode} & $\left| {RM{{\left( {{\pi ^ + }{\pi ^ - }} \right)}} - {m_{{J \mathord{\left/ {\vphantom {J \psi }} \right.  \kern-\nulldelimiterspace} \psi }}}} \right|> 10$ \\
   &   $\left| {RM{{\left( {{\pi ^ 0 }{\pi ^ 0 }} \right)}} - {m_{{J \mathord{\left/ {\vphantom {J \psi }} \right.  \kern-\nulldelimiterspace} \psi }}}} \right|>15$  \\
   &   $\left| {M\left( {{\pi ^ {0} }{\pi ^ {0} }{\pi ^{0} }} \right) - {m_{\eta} }} \right|>13$  \\

 &  $\left| {M\left( {p{\pi ^ - }} \right) - {m_\Lambda }} \right|>6$    \\
 &  $\left| {M\left( {\bar{p}{\pi ^ + }} \right) - {m_{\bar{\Lambda}} }} \right|>6$   \\

   &  \small$M\left( {p{\pi ^0}} \right) \in \left[ {1176,1197} \right]\&  M\left( {\bar p{\pi ^0}} \right) \in \left[ {1176,1197} \right]$ \\

   \hline

\multirow{2}*{\normalsize$\phi$-mode} & $\left| {RM{{\left( {{\pi ^ 0 }{\pi ^ 0 }} \right)}} - {m_{{J \mathord{\left/ {\vphantom {J \psi }} \right.  \kern-\nulldelimiterspace} \psi }}}} \right|>9$ \\
    &      \small$M\left( {p{\pi ^0}} \right) \in \left[ {1176,1200} \right]\&  M\left( {\bar p{\pi ^0}} \right) \in \left[ {1176,1200} \right]$  \\
  \hline   \hline
\end{tabular}

\end{center}
\end{table}

\begin{figure}[htbp]
\centering
  \mbox{
    \begin{overpic}[width=0.24\textwidth,clip=true]{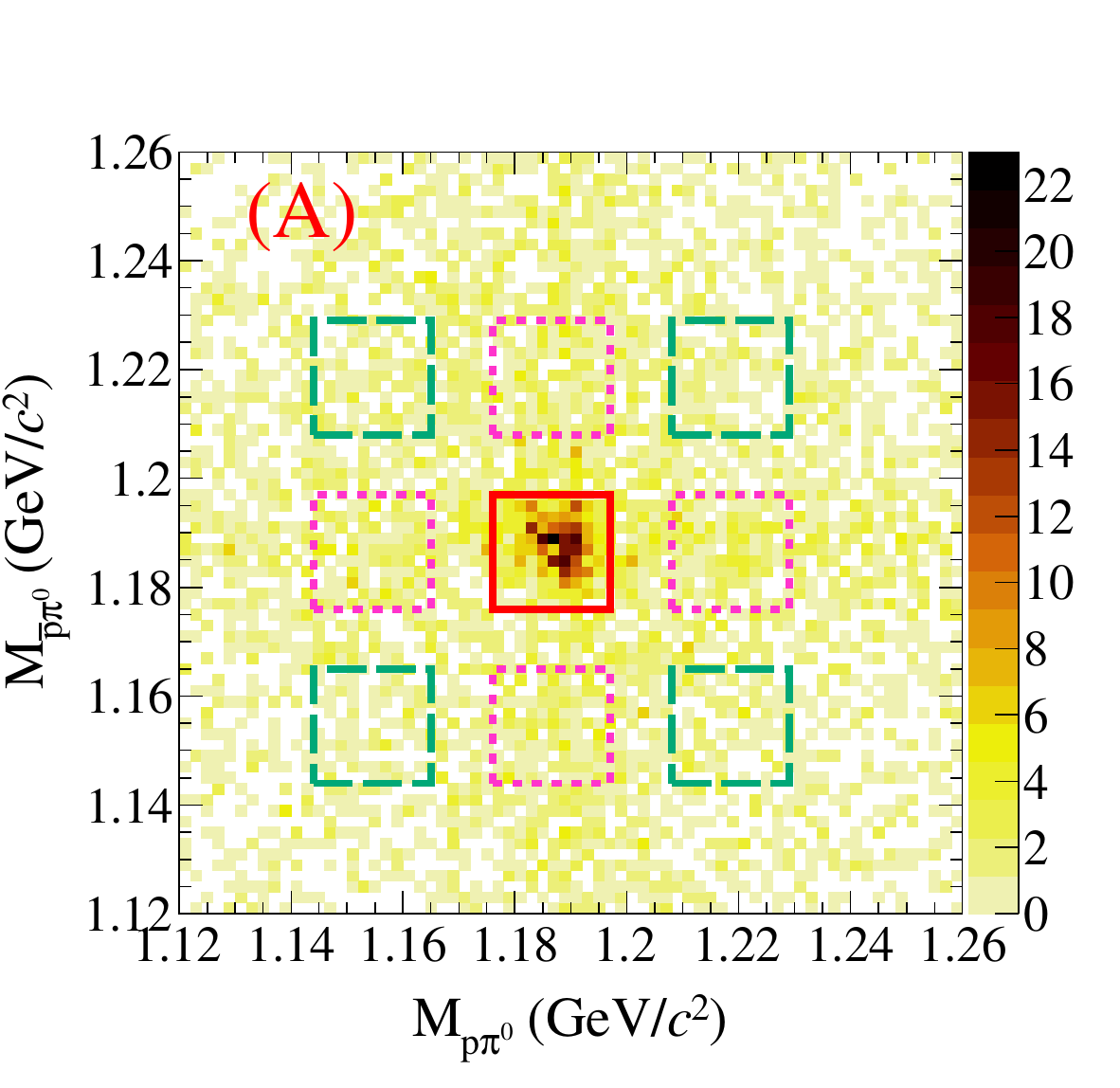}
    \end{overpic}
    \begin{overpic}[width=0.24\textwidth,clip=true]{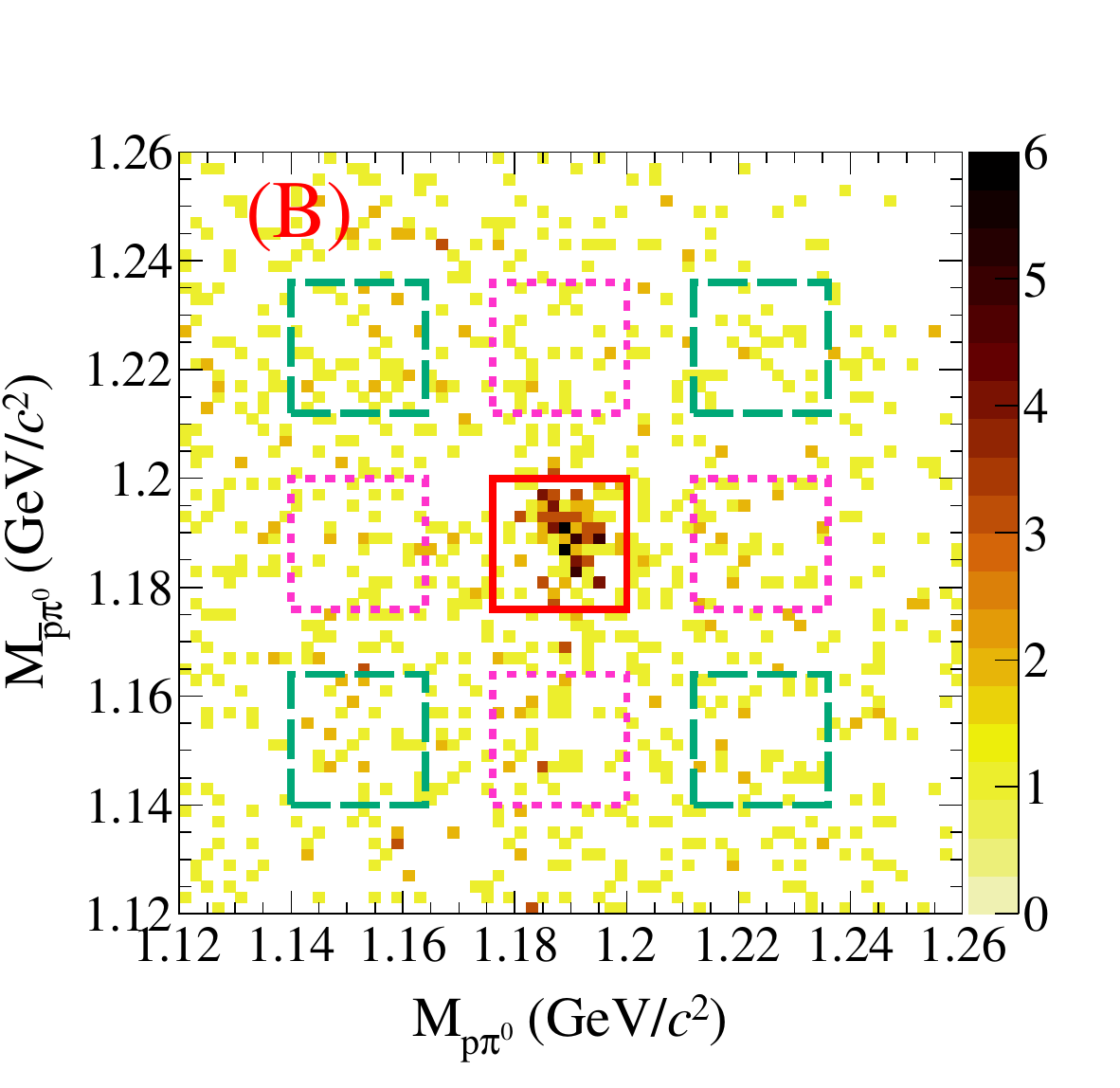}
    \end{overpic}
   }
\caption{Distributions of $M_{\bar{p}\pi^{0}}$ versus $M_{p\pi^{0}}$ of the accepted candidates for (A) $\psi(3686)\to\Sigma^{+}\bar{\Sigma}^{-}\omega$ and (B) $\psi(3686)\to\Sigma^{+}\bar{\Sigma}^{-}\phi$, where the red solid rectangle denotes the signal region, the pink dashed rectangles the sideband I region, and the green long-dashed rectangles the sideband II region.   }
\label{fig:2D_msigma}
\end{figure}

   After applying all the selection criteria, the two-dimensional (2D) distributions of the $\bar{p}\pi^{0}$ invariant mass versus the $p\pi^{0}$ invariant mass in data are shown in Fig.~\ref{fig:2D_msigma}, where clear $\SSB$ signal peaks are seen in the two decay modes. The one-dimensional (1D) $\Sigma^{+}(\bar{\Sigma}^{-})$ signal and sideband regions are defined as $\left(1176,1197\right)$ and $\left(1144,1165\right)$ or $\left(1208,1229\right)$ MeV/$c^{2}$ for the $\omega$-mode, $\left(1176,1200\right)$ and $\left(1140,1164\right)$ or $\left(1212,1236\right)$ MeV/$c^{2}$ for the $\phi$-mode, respectively. The 2D $\SSB$ signal region is defined as the square region with both $p\pi^{0}$ and $\bar{p}\pi^{0}$ combinations lying in the 1D $\Sigma^{+}(\bar{\Sigma}^{-})$ signal regions. The $\SSB$ sideband I regions are defined as the square regions with either one of the $p\pi^{0}$ or $\bar{p}\pi^{0}$ combinations located in the 1D $\Sigma^{+}(\bar{\Sigma}^{-})$ sideband regions and the other in the 1D signal region. The sideband II regions are defined as the square regions with both $p\pi^{0}$ and $\bar{p}\pi^{0}$ combinations located in the 1D $\Sigma^{+}(\bar{\Sigma}^{-})$ sideband regions.

   The potential remaining backgrounds are investigated with the inclusive $\psi(3686)$ MC samples, using the event-type analysis tool  TopoAna~\cite{zhouxy_topoAna}. According to our strategy for extracting signal yields, we consider decay processes involving $\omega$(for $\omega$-mode) or $\phi$(for $\phi$-mode) as so-called peaking backgrounds. Furthermore, peaking backgrounds were found only for the $\phi$-mode, mainly from the $\psi(3686)\to\gamma\chi_{c1,2}, \chi_{c1,2}\to\Lambda\bar{\Lambda}\phi$ processes, and contaminations from these processes were found to be negligible. For the non-peaking backgrounds, we consider them as a smooth distribution in the invariant mass distribution. The possible non-$\Sigma^{+}(\bar{\Sigma}^{-})$ and non-$\SSB$ peaking backgrounds from $\psi(3686)$ decays are studied with sideband events, and will be considered when extracting the signal yields later. We determine the corresponding peaking contributions, $\rm N_{I}^{SD}=63.0\pm14.5$ and $\rm N_{II}^{SD}=61.8\pm12.7$ for the $\omega$-mode, $\rm N_{I}^{SD}=29.5\pm8.4$ and $\rm N_{II}^{SD}=26.2\pm8.5$ for the $\phi$-mode, by fitting the data in the 2D sideband I and sideband II regions. In the fitting process, the description of the peaking contribution and the smooth background is consistent with the strategy of fitting the $\Sigma^{+}\bar{\Sigma}^{-}$ signal region events in Section~\ref{Sec:BR_determined}.

   To investigate the possible quantum electrodynamics (QED) background, the same selection criteria are applied to data samples collected at center-of-mass energies of 3.65 GeV and 3.773 GeV, corresponding to values of integrated luminosity of 454 $\rm{pb}^{-1}$ and 2931.8 $\rm{pb}^{-1}$~\cite{QED_Data_3773}, respectively. Only a peaking contribution of 1.6$\sigma$ is found for the $\omega$-mode at 3.773 GeV, and no peaking background contribution is seen for the $\phi$-mode. We also consider the system uncertainty caused by the interference effect between the resonance decay and the continuum processes, but the corresponding effect is negligible.
   \section{\label{Sec:BR_determined}Signal yields}
   The signal yields are determined by  extended unbinned maximum likelihood fits on the $\pi^{+}\pi^{-}\pi^{0}$ ($\omega$-mode) or $\kk$ ($\phi$-mode) invariant mass distributions in the $\SSB$ signal regions. The total probability density function consists of a signal  component and various background contributions. The signal component is modelled with the MC-simulated shape convolved with a Gaussian function to account for the possible difference in the mass resolution between data and MC simulation.

   For the background of the 2D sideband I region and sideband II region, the shape is described using the MC-simulated shape of the signal, while the number of events is fixed to a normalization value, i.e. $\rm{\frac{1}{2}}\times {\rm{N_{I}^{SD}}}-\rm{\frac{1}{4}}\times {\rm{N_{II}^{SD}}}$.  According to the calculation in section~\ref{Sec:Selection}, the background contributions from all 2D sideband regions are fixed as $16.1$ events for the $\omega$-mode, and $8.2$ events for the $\phi$-mode.

\begin{figure}[htbp]

		\begin{minipage}[t]{0.82\linewidth}
		\includegraphics[width=1\textwidth]{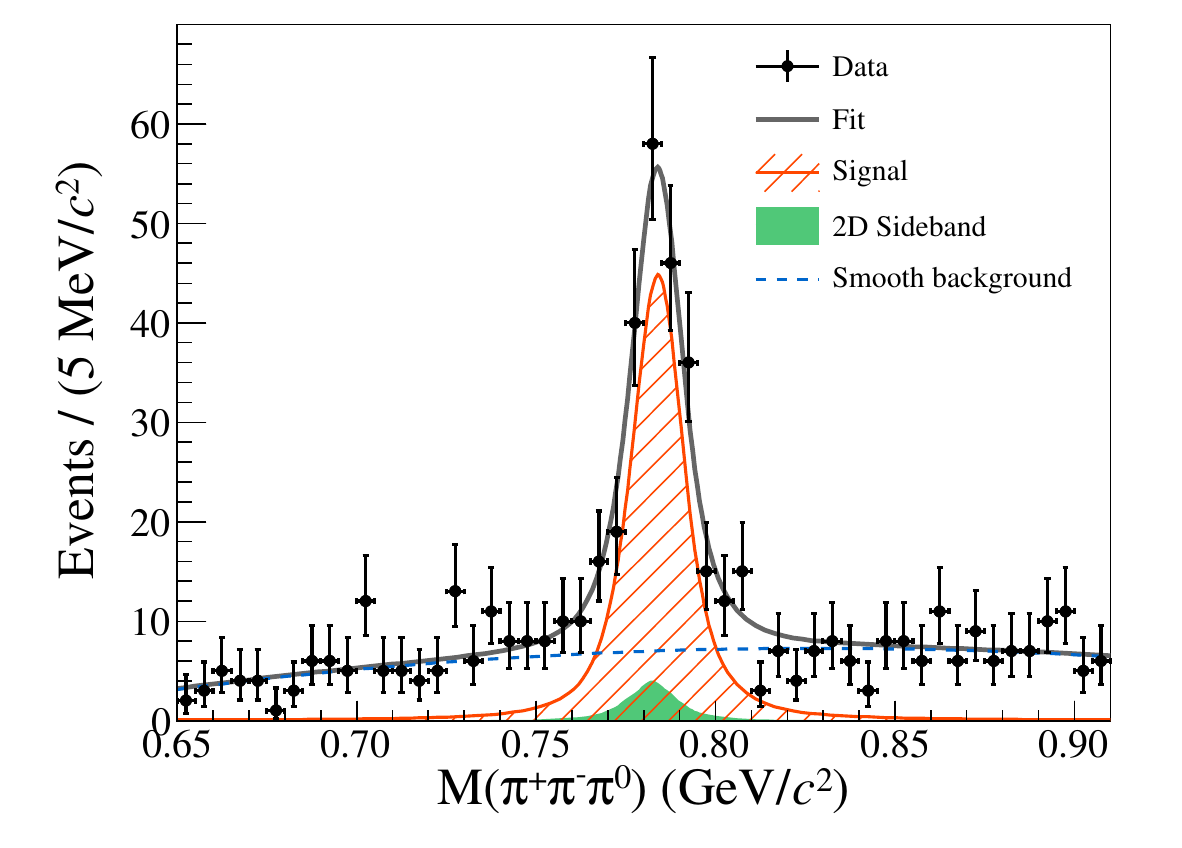}
		\end{minipage}

	\caption{The $M(\pi^{+}\pi^{-}\pi^{0})$ distribution of the accepted events in the 2D signal region. The black points with uncertainties are data, the gray solid curve is the fit result, the orange shaded histogram represents the signal, the green histogram denotes the scaled 2D sideband contribution and the blue dotted curve represents the remaining smooth background.}
		\label{fig:momg}
		\end{figure}

\begin{figure}[htbp]
		\begin{minipage}[t]{0.82\linewidth}
		\includegraphics[width=1\textwidth]{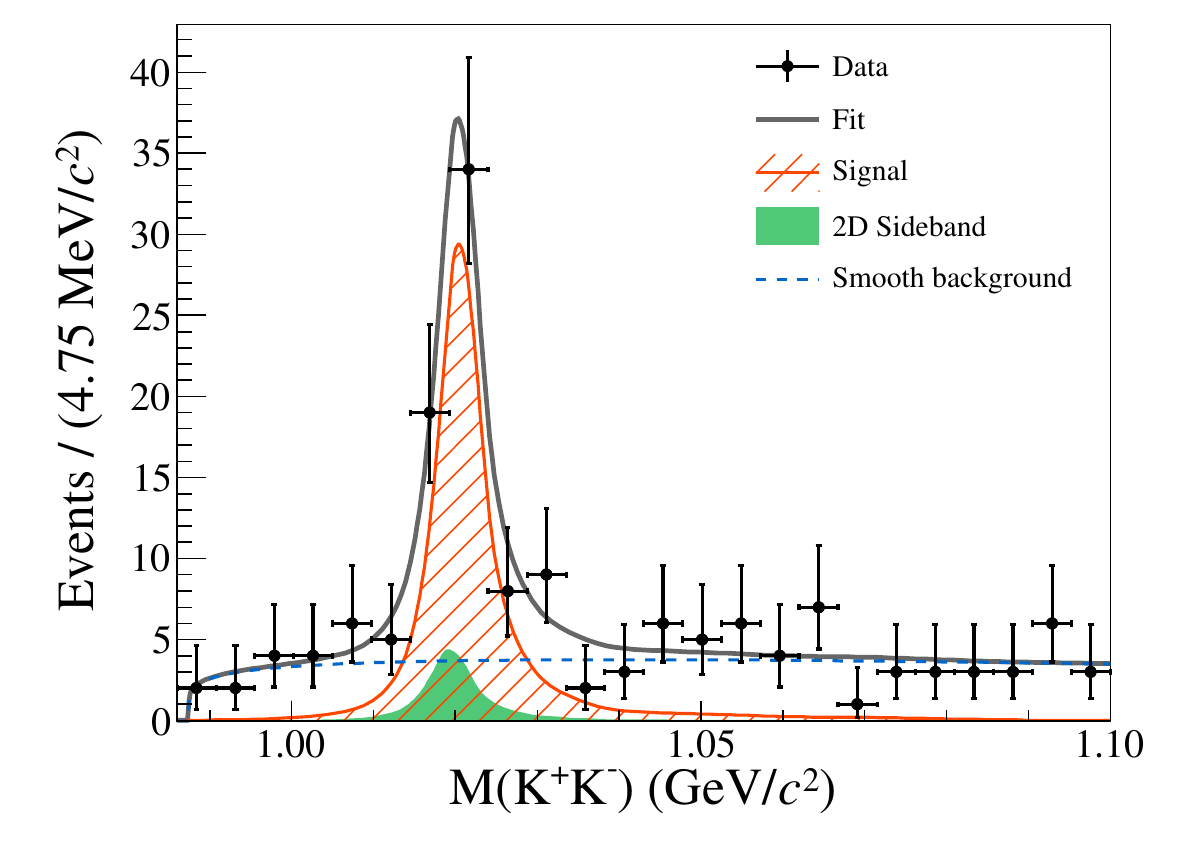}
		
		\end{minipage}
		\caption{The $M(K^{+}K^{-})$ distribution of the accepted events in the 2D signal region. The black points with uncertainties are data, the gray solid curve is the fit result, the orange shaded histogram represents the signal, the green histogram denotes the scaled 2D sideband contribution and the blue dotted curve represents the remaining smooth background.}
		\label{fig:mkk}
		\end{figure}

   The remaining smooth backgrounds for the two modes are modelled by a second-order polynomial function and an Argus function, respectively. The numbers of fitted signal events are $198.7\pm18.9$ and $55.2\pm10.0$ for $\omega$-mode and $\phi$-mode, respectively. The corresponding statistical significances are determined to be 13.8$\sigma$ and $7.6\sigma$ for $\omega$-mode and $\phi$-mode, respectively. The statistical significance is estimated by the likelihood difference between the fits with and without the signal component, taking the change in the degrees of freedom into account. Figures~\ref{fig:momg} and~\ref{fig:mkk} show the fit results.

\begin{figure*}[htbp]
\begin{center}

\begin{minipage}[t]{0.28\linewidth}
\includegraphics[width=1\textwidth]{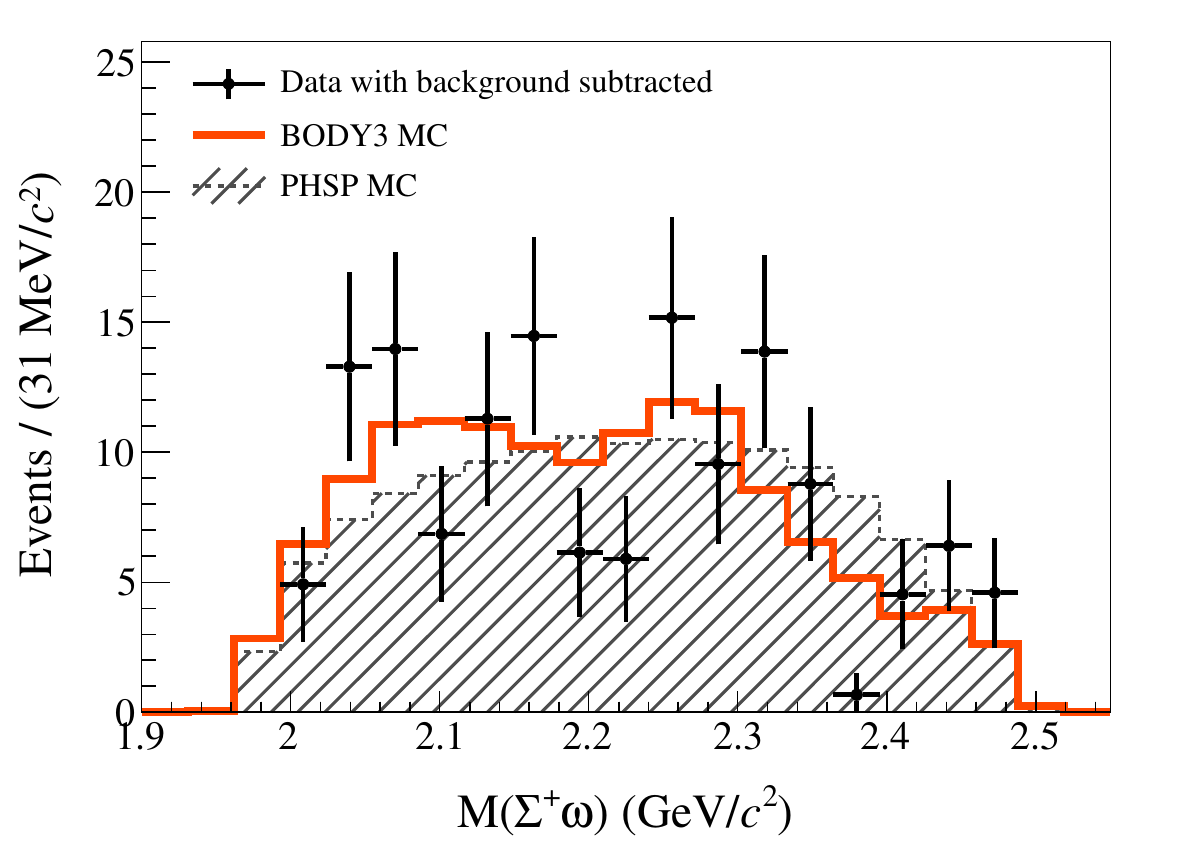}
\end{minipage}
\begin{minipage}[t]{0.28\linewidth}
\includegraphics[width=1\textwidth]{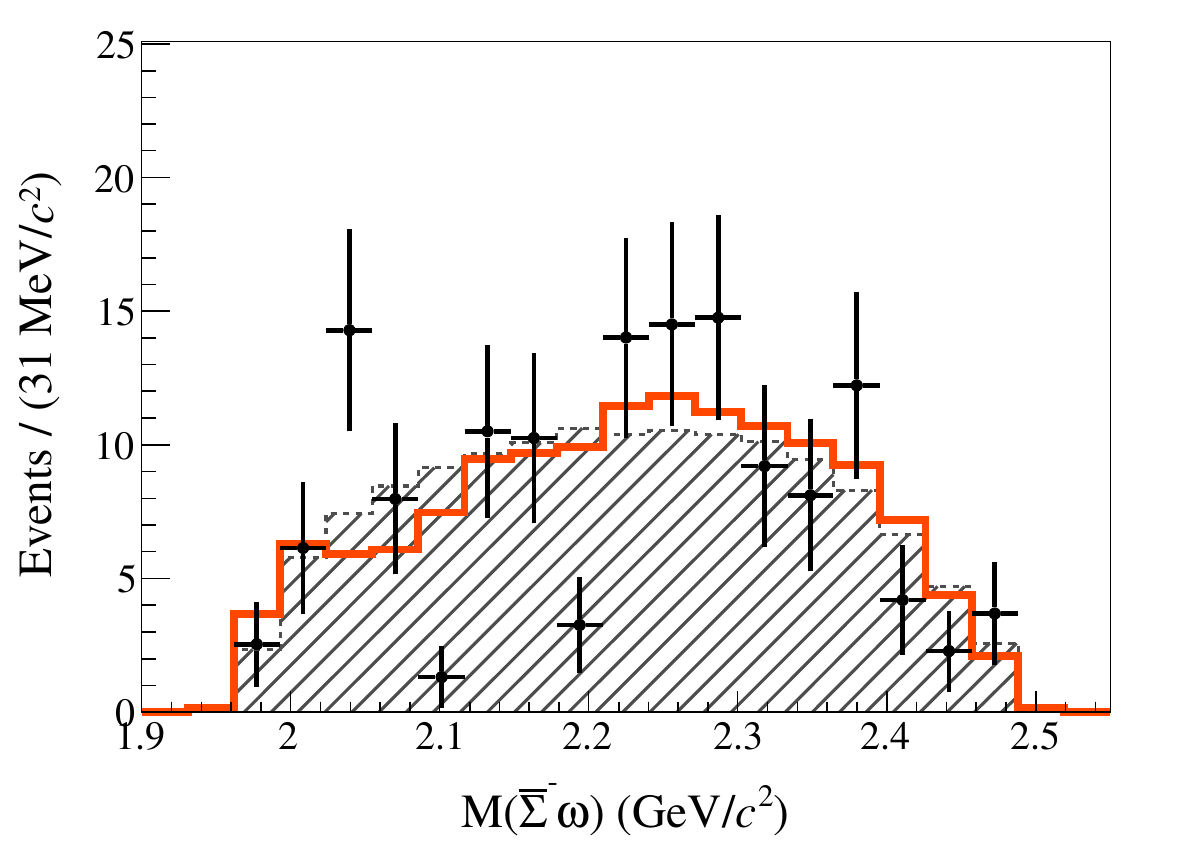}
\end{minipage}
\begin{minipage}[t]{0.28\linewidth}
\includegraphics[width=1\textwidth]{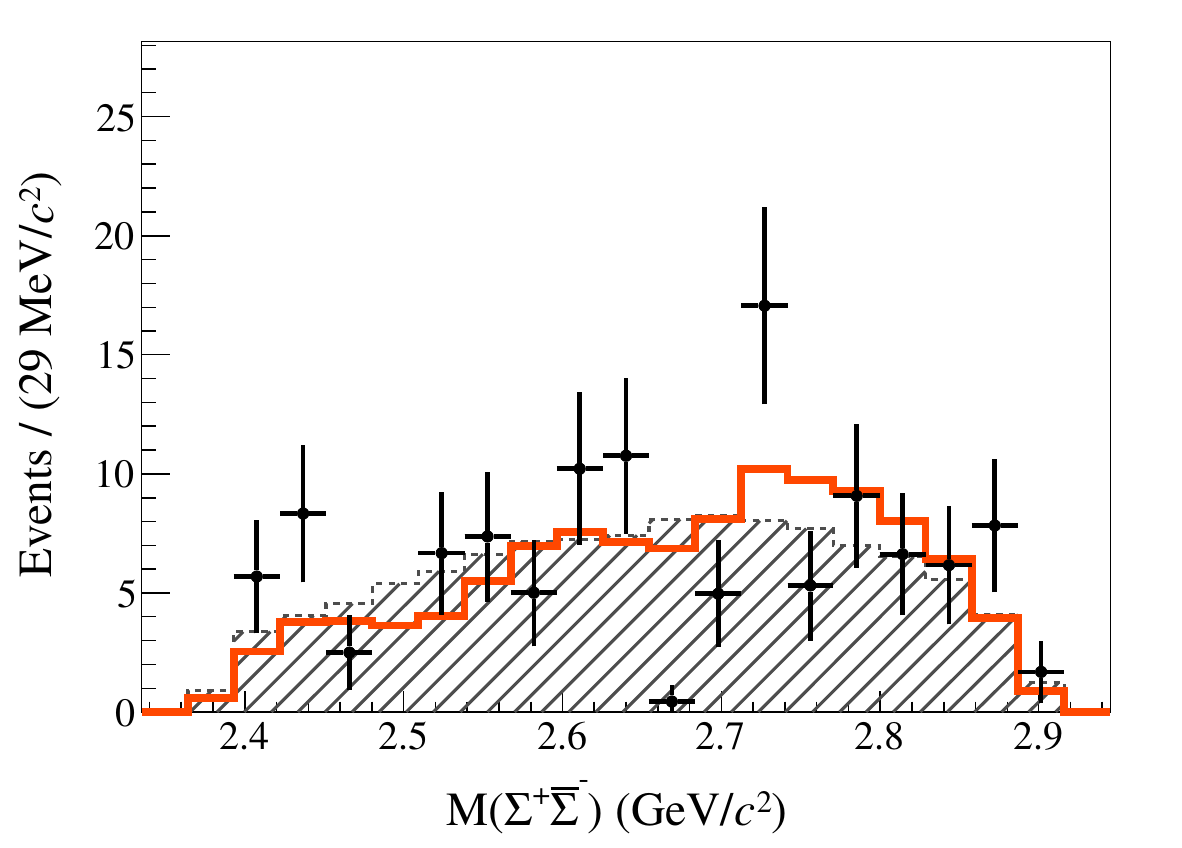}
\end{minipage}%

\begin{minipage}[t]{0.28\linewidth}
\includegraphics[width=1\textwidth]{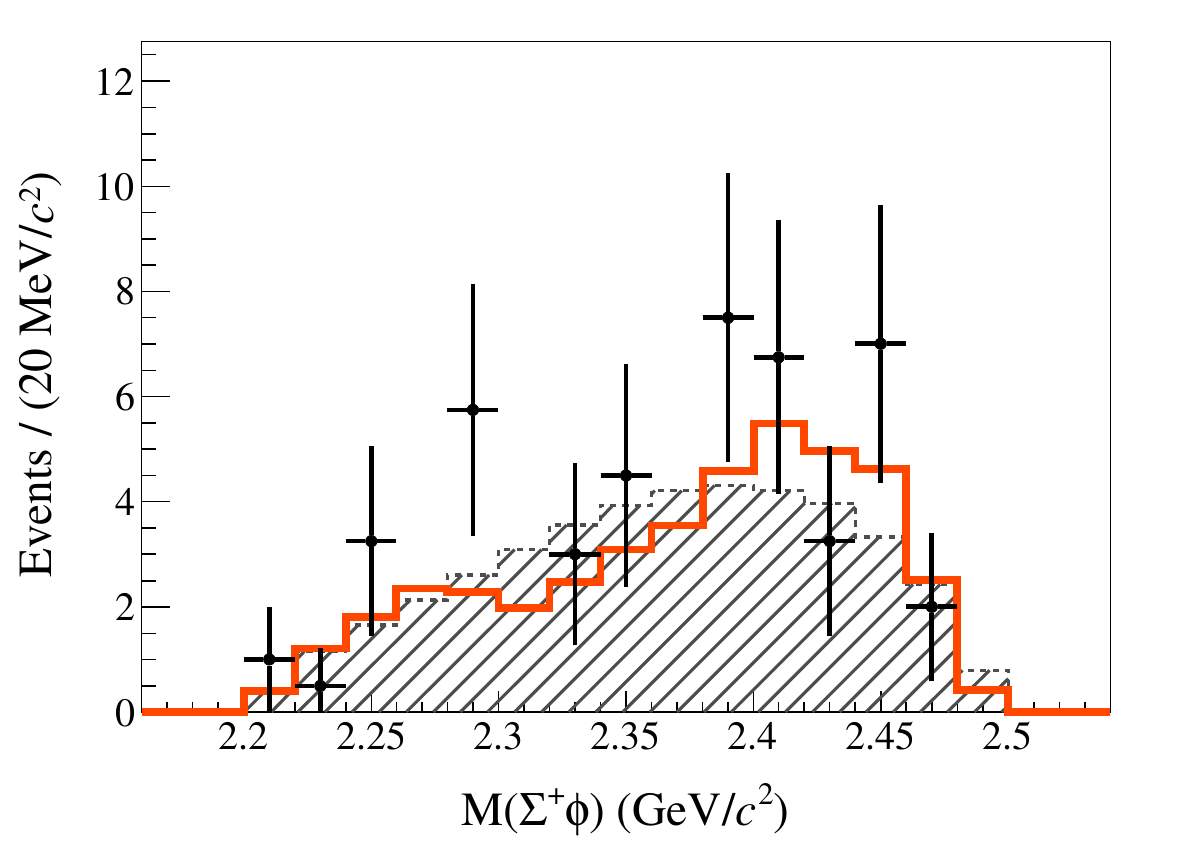}
\end{minipage}
\begin{minipage}[t]{0.28\linewidth}
\includegraphics[width=1\textwidth]{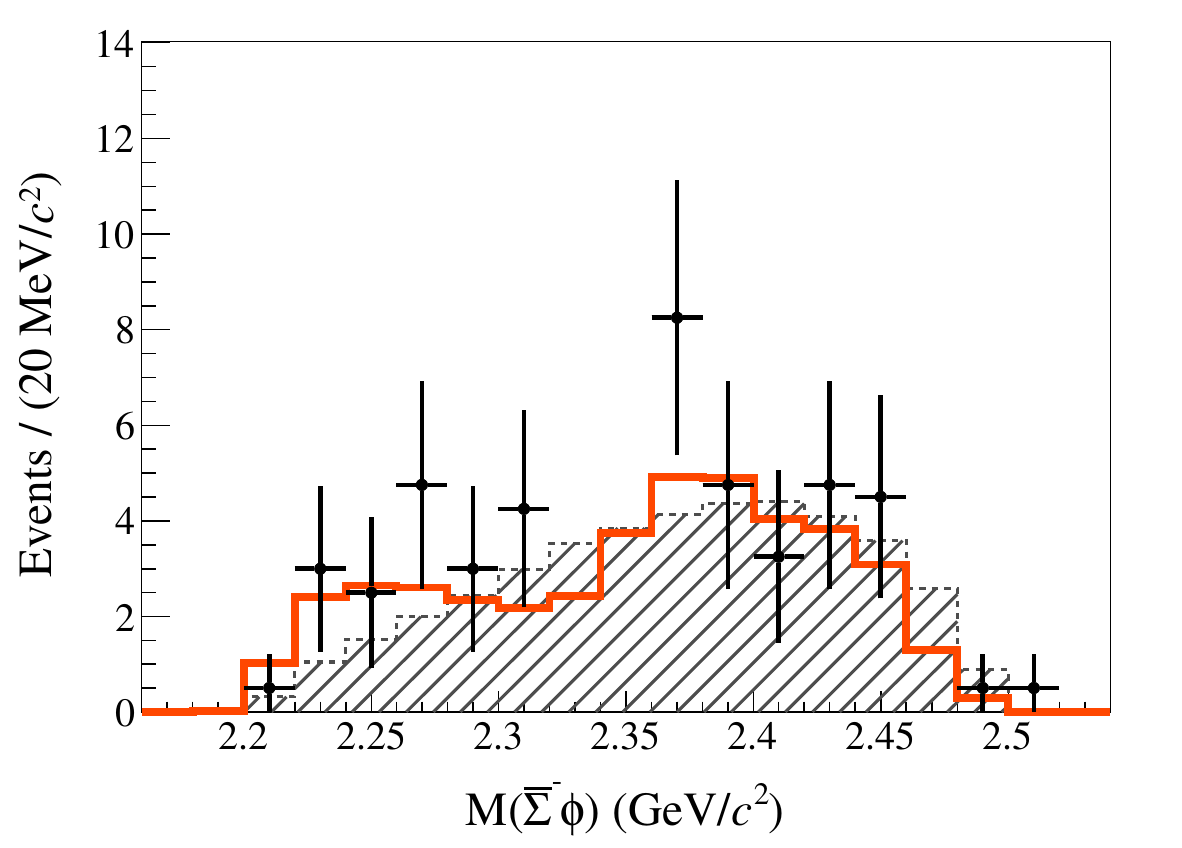}
\end{minipage}
\begin{minipage}[t]{0.28\linewidth}
\includegraphics[width=1\textwidth]{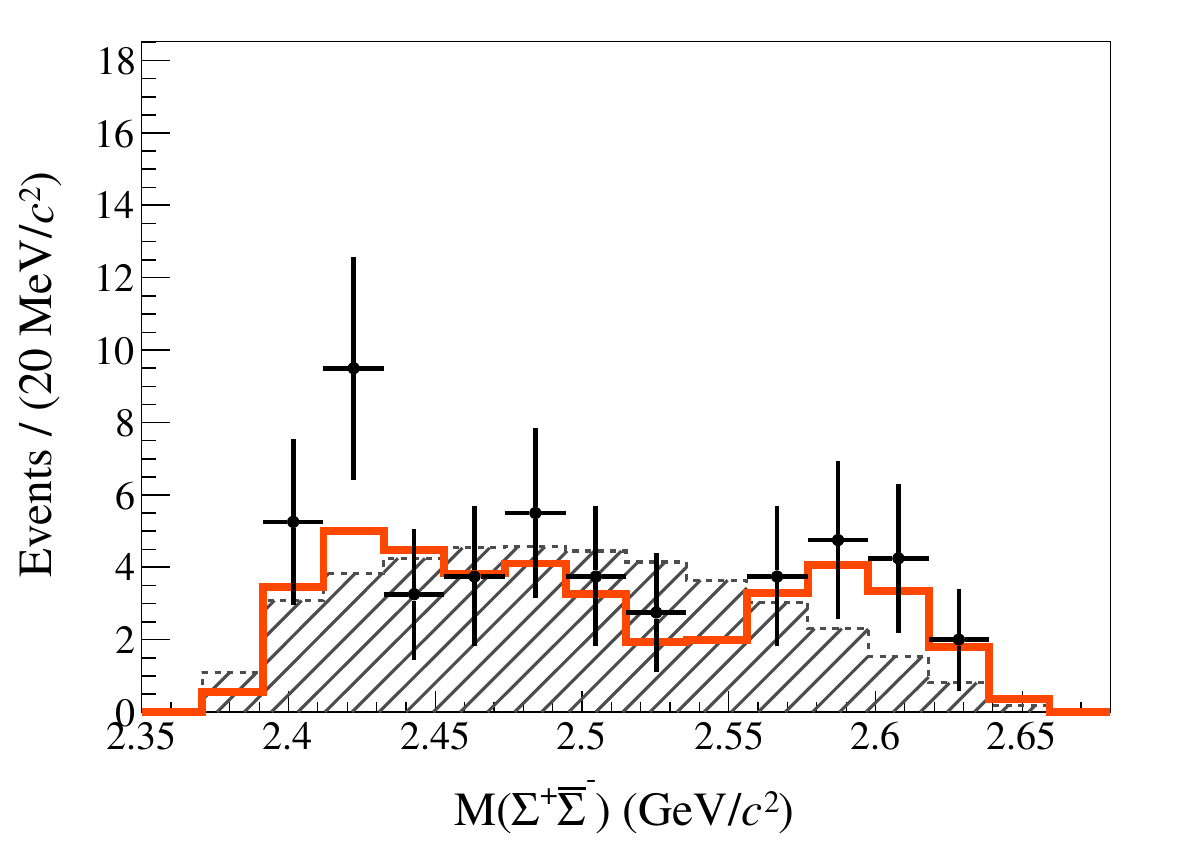}
\end{minipage}%

\caption{Invariant mass distributions of all the two-body particle combinations for (upper row) $\psi(3686)\to\Sigma^{+}\bar{\Sigma}^{-}\omega$ and (lower row)  $\psi(3686)\to\Sigma^{+}\bar{\Sigma}^{-}\phi$. The points with error bars represent the background-subtracted data, the orange curves and the grey hatched histograms are the BODY3 and PHSP signal MC events, respectively.}
\label{BODY3_hists_SSV}
\end{center}
\end{figure*}

The potential intermediate states are investigated via Dalitz plots and their 1D projections. Figure~\ref{BODY3_hists_SSV} shows the invariant mass distributions of the different two-body combinations for the two decay modes, where the background contributions have been subtracted by using the normalized sideband events. No obvious structure can be found in any of the mass distributions. Nevertheless, the experimental distributions are not consistent with the signal MC sample generated according to the phase space (PHSP) distribution. To improve the accuracy of the detection efficiency, the PHSP model is replaced by the modified data-driven generator BODY3, where the MC-simulated events are sampled according to the Dalitz distribution in the data. The comparisons are shown in Fig.~\ref{BODY3_hists_SSV}. The MC sample of BODY3 describes the data better.

\section{Results for branching fractions}
The BF for each signal decay is calculated by:
\begin{equation}\label{determine_branch}
\mathcal{B}\left( {{\psi(3686)} \to \Sigma^{+}\bar{\Sigma}^{-}{\rm{V}}} \right) = {\textstyle{{{N_{obs}}} \over { {N_{\psi \left( {3686} \right) }}\cdot {\prod _{i}}  \mathcal{{B}}_{i} \cdot \varepsilon }}}.
      \end{equation}
Here, $N_{obs}$ is the number of signal events determined by the fit, $N_{\psi(3686)}$ is the total number of $\psi(3686)$ events, $\rm{V}$ denotes the vector meson $\omega(\phi)$, $\mathcal{{B}}_{i}$ is the BF of the $i$-th intermediate state taken from the PDG~\cite{pdg2022},
$\varepsilon$ is the detection efficiency, which is determined by the MC simulation based on the BODY3 model. The corresponding numerical values are listed in Table~\ref{list_summary}.

\begin {table*}[htbp]
\begin{center}
\renewcommand\arraystretch{1.2}

    {\caption {Signal yields ($N_{obs}$), corrected signal efficiencies ($\varepsilon$), BFs ($\Br$) and statistical significances $(\sigma )$ for the two decay modes.}
    \label{list_summary}}
    \begin {tabular}{c c  c}\hline\hline
	        &   $\psi(3686)\to\Sigma^{+}\bar{\Sigma}^{-}\omega$      &  $\psi(3686)\to\Sigma^{+}\bar{\Sigma}^{-}\phi$ \\  \hline

  $N_{obs}$      &  $198.7\pm18.9$       &   $55.2\pm10.0$ \\	
  $\varepsilon$ (\%)  &          1.69              &     5.33               \\
	   $\Br$   &      $\left( { 1.90 \pm 0.18 \pm 0.21 }  \right) \times {10^{ - 5}}$  & $\left( { 2.96 \pm 0.54 \pm 0.41 }  \right) \times {10^{ - 6}}$   \\
$\sigma$ &     13.8   &      7.6  \\

\hline
\hline
\end{tabular}

\end{center}
\end{table*}

\section{Systematic Uncertainties}\label{sec:sysU}
As sources of systematic uncertainties in the BF measurements we consider the tracking and PID efficiencies, the photon detection efficiency, the $\pi^{0}$ reconstruction efficiency, the kinematic fit, the procedure for signal extraction and background subtraction, the MC simulation modeling, and so on. They are described as follows.

\begin{itemize}

\item[(i)]{{\bf Tracking and PID:}}
The uncertainties due to the tracking efficiency are estimated with the control samples $\psi(3686)\to\pi^{+}\pi^{-}J/\psi$, $J/\psi \rightarrow K^0_S K^\pm \pi^\mp$, and $J/\psi \to p\bar{p}\pi^{+}\pi^{-}$, and are determined to be 1\%~\cite{Ablikim:092009,Ablikim:2011kv,error_ppb2pi} for each charged track. The PID uncertainties are determined to be 1\% for each charged track as well, based on the same samples used to estimate the tracking uncertainties.

\item[(ii)]{\bf Photon detection:} The difference in the photon detection efficiencies between data and MC simulation is studied using a control sample of $e^{+}e^{-}\to\gamma\mu^{+}\mu^{-}$ and found to be less than 0.5\% for each photon.

\item[(iii)]{\bf $\pi^{0}$ reconstruction}:
Based on the control samples $\psi(3686)\to\pi^{0}\pi^{0}J/\psi,J/\psi\to l^{+}l^{-}$ and $e^{+}e^{-}\to\omega\pi^{0}$ at $\sqrt{s}=3.773$ GeV, the relative difference of the $\pi^{0}$ reconstruction efficiencies between data and MC simulation has been estimated, and the results on the two datasets are consistent with each other. We have studied the relative difference as a function of the polar angle and of the total $\pi^{0}$ momentum $p$, and it decreases linearly as a function of the momentum following $(0.06-2.41\times p)\%$. The resulting systematic uncertainties of the $\pi^{0}$ reconstruction efficiency are determined to be 2.9\% and 0.8\% for the $\omega$-mode and the $\phi$-mode, respectively.

\item[(iv)]{\bf Kinematic fit}:
The uncertainty associated to the kinematic fit is estimated by comparing the efficiencies with and without the helix parameter correction~\cite{YPG:bam}.

\item[(v)]{\bf Signal yields}
  \begin{itemize}
  \item[\textbullet] {\bf Mass window}:
   The systematic uncertainties related to each individual mass window requirement for background rejection are estimated by varying the size of each mass window by one standard deviation of the corresponding mass resolution. The larger variation in the BF for each mass requirement is considered as the related systematic uncertainty. For the 2D mass window of $\Sigma^{+}\bar{\Sigma}^{-}$ signal selection, the uncertainty is estimated by the control sample $J/\psi\to\Sigma^{+}\bar{\Sigma}^{-}$. The differences in the selection efficiencies between data and MC simulation from the control sample are taken as the corresponding systematic uncertainties.

  \item[\textbullet]  {\bf Fitting range}:
   The uncertainty due to the fitting range is estimated by randomly changing the range of the fits (for the $\omega$-mode, the lower and the upper side in the range [630, 670] MeV/$c^2$ and [890, 930] MeV/$c^2$, respectively; for the  $\phi$-mode, the upper side in the range [1080, 1120] MeV/$c^2$) and performing the fit 800 times, using a recalculated signal efficiency, as described in Ref.~\cite{maxuning}. The variance of the re-measured BFs is taken as the corresponding systematic uncertainty, respectively.

  \item[\textbullet]  {\bf Signal shape}:
   The uncertainty due to the signal shape is estimated by replacing the MC-simulated shape convolved with a Gaussian function with only the MC-simulated shape. The difference in the measured BF is taken as the systematic uncertainty.

  \item[\textbullet]  {\bf Non-peaking background}:
   To estimate the uncertainty of the non-peaking background in the fit, we perform alternative fits by replacing the second-order Chebychev polynomial with a third-order polynomial function or the Argus function with the third-order Chebychev polynomial for the data. The difference in the measured BF is taken as the systematic uncertainty.

   \item[\textbullet]  {\bf $\Delta$-related background}:
    The uncertainty due to the possible peaking background $\psi(3686)\to\Delta^{+}\bar{\Delta}^{-}\omega(\phi)$ is estimated by including in the fit this channel, assuming $\Br(\Delta^+\to p\pi^{0})=66.6\%$ and $\Br(\psi(3686)\to\Delta^{+}\Delta^{-}\omega(\phi))$ equal to $\Br(\psi(3686)\to\Sigma^{+}\Sigma^{-}\omega(\phi))$, and the largest differences between these results and the nominal ones are taken as the systematic uncertainty, separately for the $\omega$-mode and the $\phi$-mode.

    \item[\textbullet]  {\bf Sideband background}:
    The systematic uncertainty associated to the 2D sideband background is estimated by increasing and reducing the sideband regions by one standard deviation of the mass resolution, or change the normalized value of the background contribution by one standard deviation of the statistical uncertainty. The difference in the measured BF is taken as the systematic uncertainty.

 \item[\textbullet] {\bf Interference}: The effect due to the interference between the $\psi(3686)$ decay and the continuum production is estimated with the method described in Ref.~\cite{wangting}. The cross section for a certain exclusive final state $f$ can be written as
 \beq \centering
\sigma _{tot}^f\left( s \right) = {\left| {a_c^f\left( s \right) + {e^{i\Phi }} \cdot a_R^f\left( s \right)} \right|^2},
\eeq
where $\sqrt{s}$ is the center-of-mass energy, $a_{c}^{f}$ and $a_{R}^{f}$ are the amplitudes of the continuum contribution and of the $\psi(3686)\to\Sigma^{+}\bar{\Sigma}^{-}\omega$ decay, respectively, and $\Phi$ is the relative phase between the two amplitudes~\cite{inter_ycz}. The systematic uncertainty due to the interference effect is assigned as the difference of $r_{R}^{f}$ between the constructive ($\Phi=\pi/2$) and destructive ($\Phi=3\pi/2$) interference, where $r_{R}^{f}$ represents the ratio of cross section from the interference term with respect to the cross section of $\psi(3686)\to\Sigma^{+}\bar{\Sigma}^{-}\omega$. Since the prior distribution of the interference angle is assumed to be uniform, this difference divided by $\sqrt{12}$, namely 3.1\%, is assigned as the corresponding systematic uncertainty for the $\omega$-mode. As previously mentioned, this uncertainty for the $\phi$-mode is found negligible.
     \end{itemize}

\item[(vi)]{\bf MC simulation}:
The systematic uncertainty associated to the MC simulation is related to the use  of the BODY3 generator and to the differences in the polar angle distributions of each final state particle between data and MC simulation. The first contribution is estimated by varying the bin size of the input Dalitz plot by $\pm25\%$, shifting the 1D sideband regions to the left or to the right by one standard deviation of the mass resolution, and changing in the BODY3 generator the background level in the input Dalitz plot  by  $\pm1\sigma$, where $\sigma$ denotes the statistical uncertainty of the background level determined from the fit result. Combining the results from the three sources, the largest change to the nominal detection efficiency is taken as the systematic uncertainty. The second  contribution is estimated by rescaling the corresponding distribution from the MC simulation to match the data. The difference in efficiency before and after the rescaling is taken as the systematic uncertainty.

\item[(vii)]{\bf Quoted BFs}:
The uncertainties for the quoted BFs are taken from PDG~\cite{pdg2022}.

\item[(viii)]{\bf $N_{\psi(3686)}$}:
The uncertainty due to the total number of $\psi(3686)$ events,
determined with inclusive hadronic $\psi(3686)$ decays, is 0.5\%~\cite{psip_num_0912}.

\end{itemize}

\begin {table}[htbp]

\renewcommand\arraystretch{1.2}
{\caption {The systematic uncertainties for $\psi(3686)\to\Sigma^{+}\bar{\Sigma}^{-}$V(V$=\omega,\phi$) decay channels (in percent).}
\label{list_sys}}
\begin {tabular}{l c c  c}\hline\hline

Source & $\Sigma^{+}\bar{\Sigma}^{-}\omega$ & $\Sigma^{+}\bar{\Sigma}^{-}\phi$    \\   \hline
	
 Tracking     &     2.9      &  2.9                                          \\
 	    PID          &     2.9     &  2.9 	                                 \\
	Photon       &     $3.0$     &  $1.0$                                          \\
	
	$\pi^{0}$ reconstruction  & $2.9$ & $0.8$   \\
	Kinematic fit  &     $1.5$       &  $1.9$                                    \\
	Mass window       &    3.0     & 2.0                                 \\
    Fitting range	&  $0.9$     &  $1.0$                                   \\
    Signal shape	&  $0.8$      &  $1.8$                                  \\
    Non-peaking background&  1.6 & $2.2$                                    \\
    $\Delta$-related background &   negligible   &  $0.2$     \\

Sideband background & 4.5   &   9.4     \\
    Interference & 3.1 & negligible \\

MC simulation & 6.6  &  8.1 \\

Quoted BFs & 1.4      & 1.6      \\

$N_{\psi(3686)}$  &     0.5      &        0.5   \\  \hline
	 Sum & 11.2 &  13.9         \\

\hline
\hline
\end{tabular}

\end{table}

     All the values of the systematic uncertainties are summarized in Table~\ref{list_sys}, for the $\psi(3686)\to\Sigma^{+}\bar{\Sigma}^{-}\omega$ and $\psi(3686)\to\Sigma^{+}\bar{\Sigma}^{-}\phi$ decays, where the total uncertainties are given by the quadratic sum, assuming statistical independence of all the contributions.

	\section{Summary}\label{sec:summary}
   Using $(27.08\pm0.14)\times10^{8}$ $\psi(3686)$ events collected with the BESIII detector at the BEPCII collider, we observe the decay modes $\psi(3686)\to\Sigma^{+}\bar{\Sigma}^{-}$V(V$=\omega,\phi$) for the first time. The corresponding branching fractions are measured to be $\Br\left(\psi(3686) \rightarrow \Sigma^{+}\bar{\Sigma}^{-} \omega\right)=(1.90~\pm~0.18~\pm~0.21)\times 10^{-5}$ and $\Br\left(\psi(3686) \rightarrow \Sigma^{+}\bar{\Sigma}^{-} \phi\right)=(2.96~\pm~0.54~\pm~0.41)\times 10^{-6}$, where the first uncertainties are statistical and the second systematic. The results are comparable to those of the previously observed decays $J/\psi(\psi(3686))\to \Sigma^{+}\bar{\Sigma}^{-}\eta$~\cite{xiaohao_SSP}, $\psi(3686)\to \Lambda\bar{\Lambda}\pi^{0}(\eta)$~\cite{wangshi_LLP} and $\psi(3686)\to \Lambda\bar{\Lambda}\omega$~\cite{zhh_LLomg}. No evident structures deviating from 3-body phase-space distributions have been observed in the invariant mass distributions of $\Sigma^{+}\bar{\Sigma}^{-}$ or $\Sigma\omega(\phi)$ pairs.
	\acknowledgements

The BESIII Collaboration thanks the staff of BEPCII and the IHEP computing center for their strong support. This work is supported in part by National Key R\&D Program of China under Contracts Nos. 2020YFA0406300, 2020YFA0406400; National Natural Science Foundation of China (NSFC) under Contracts Nos. 11635010, 11735014, 11835012, 11935015, 11935016, 11935018, 11961141012, 12022510, 12025502, 12035009, 12035013, 12061131003, 12192260, 12192261, 12192262, 12192263, 12192264, 12192265, 12221005, 12225509, 12235017, 12150004; Program of Science and Technology Development Plan of Jilin Province of China under Contract No. 20210508047RQ and 20230101021JC; the Chinese Academy of Sciences (CAS) Large-Scale Scientific Facility Program; the CAS Center for Excellence in Particle Physics (CCEPP); Joint Large-Scale Scientific Facility Funds of the NSFC and CAS under Contract No. U1832207; CAS Key Research Program of Frontier Sciences under Contracts Nos. QYZDJ-SSW-SLH003, QYZDJ-SSW-SLH040; 100 Talents Program of CAS; The Institute of Nuclear and Particle Physics (INPAC) and Shanghai Key Laboratory for Particle Physics and Cosmology; European Union's Horizon 2020 research and innovation programme under Marie Sklodowska-Curie grant agreement under Contract No. 894790; German Research Foundation DFG under Contracts Nos. 455635585, Collaborative Research Center CRC 1044, FOR5327, GRK 2149; Istituto Nazionale di Fisica Nucleare, Italy; Ministry of Development of Turkey under Contract No. DPT2006K-120470; National Research Foundation of Korea under Contract No. NRF-2022R1A2C1092335; National Science and Technology fund of Mongolia; National Science Research and Innovation Fund (NSRF) via the Program Management Unit for Human Resources \& Institutional Development, Research and Innovation of Thailand under Contract No. B16F640076; Polish National Science Centre under Contract No. 2019/35/O/ST2/02907; The Swedish Research Council; U. S. Department of Energy under Contract No. DE-FG02-05ER41374.


\end{document}